\documentclass[iop,appendixfloats,revtex4]{emulateapj}
\usepackage{natbib}
\usepackage{hyperref}
\usepackage{amsmath}
\usepackage{graphicx}
\usepackage{verbatim}
\usepackage{epsfig}
\usepackage{xcolor}
\usepackage{ulem}
\usepackage{gensymb}
\usepackage{textcomp}
\usepackage{booktabs}
\usepackage{commath}

\shorttitle{Binding energy of molecules on ice}
\shortauthors{He et al.}
\begin{document}

\title{Binding Energy of Molecules on Water Ice: Laboratory Measurements and Modeling}

\author{Jiao He\altaffilmark{1}, Kinsuk Acharyya\altaffilmark{2}, \& Gianfranco Vidali\altaffilmark{1}}

\affil{$^1$Physics Department, Syracuse University, Syracuse, NY 13244, USA}
\affil{$^2$PLANEX, Physical Research Laboratory, Ahmedabad, 380009, India}
\email{gvidali@syr.edu}

\begin{abstract}
We measured the binding energy of N$_2$, CO, O$_2$, CH$_4$, and CO$_2$ on non-porous (compact) amorphous solid water (np-ASW),  of N$_2$ and CO on porous amorphous solid water (p-ASW), and of NH$_3$ on crystalline water ice. We were able to measure binding energies down to a fraction of 1\% of a layer, thus making these measurements more appropriate for astrochemistry than the existing values. We found that CO$_2$ forms clusters on np-ASW surface even at very low coverages. The binding energies of N$_2$, CO, O$_2$, and CH$_4$ decrease with coverage in the submonolayer regime. Their values at the low coverage limit are much higher than what is commonly used in gas-grain models. An empirical formula was used to describe the coverage dependence of the binding energies. We used the newly determined binding energy distributions in a simulation of gas-grain chemistry for cold cloud and hot core models. We found that owing to the higher value of desorption energy in the sub-monlayer regime a fraction of all these ices stays much longer and up to higher temperature on the grain surface compared to the single value energies currently used in the astrochemical models.
\end{abstract}

\keywords{ISM: molecules---ISM: atoms---ISM: abundances---ISM:dust, extinction---Physical Data and Processes: astrochemistry}

\section{Introduction}

In the last two decades, the importance of considering the chemistry occurring on grains or with grains
has become clear \citep{Vidali2013,Linnartz2015}, thanks to an ever increasing number of laboratory experiments required
to explain new observational data on ices coating dust grains \citep{Boogert2015}. Simulations of ISM
environments make now use of gas-grain chemical networks, whether they employ rate-equations \citep{Garrod2008},
kinetic Monte Carlo \citep{Vasyunin2009}, or other stochastic methods \citep{Biham2001}. One key parameter
that influences the ISM chemistry is the binding energy of a particle on a dust grain. The residence time of
a particle on a grain is directly related to its binding energy to the grain through $t=\tau
\exp(E_{\rm D}/k_{\rm B}T_{\rm s})$, where $E_{\rm D}$ is the binding energy, $T_{\rm s}$ is the temperature of the surface and
$\tau$ is a characteristic time often associated, in the simplest case, to the inverse vibrational frequency
of the particle in the adsorption well. Furthermore, in models the energy barrier for thermal diffusion is estimated by an empirical relation, $R=E_{\rm d}/E_{\rm D}$, where E$_{\rm d}$ is the energy barrier to diffusion. For weakly held particles on surfaces---the ones of interest here, $R\sim0.3$ for crystalline surfaces \citep{Antczak2005,Bruch2007}, but higher (0.5--0.8) for disordered/amorphous surfaces
\citep{Katz1999,Garrod2008}.  Modern day astrochemical models used a fixed binding energy for desorption;
however, in reality the binding energy is a function of coverage. Recently, we found an empirical correlation
between the sticking coefficient (or the probability for a particle to stick to a surface) and the binding
energy for simple molecules on non-porous amorphous solid water (np-ASW) \citep{He2016a}. To know when and
how molecules desorb from ices is obviously very important, especially in hot cores scenarios, where photons
from the new star warm up the surrounding ice-coated grains causing an increase in detectable abundance of
molecules in the gas-phase \citep{Ikeda2001,Bisschop2007}.

To obtain the binding energy of atoms or molecules on a surface, the technique of Thermal Programmed Desorption
(TPD) is often used; see \citet{Kolasinski2008} and next section for details. Although TPD measures the desorption
energy, or the energy necessary to desorb a particle, this energy is essentially the binding energy if there are
no activated processes, as it should be the case for simple molecules on ices. While there have been numerous
laboratory experiments for determining the binding energy of molecules on ices \citep{Collings2004,Burke2010,
Noble2012,Martin2014,Collings2015,Smith2016}, most of them were done at monolayer or multilayer coverage.
However, when disordered or amorphous surfaces are concerned, which is the case of ices in the ISM, there is
a notable variation of the binding energy with coverage, as in the case of D$_2$ on porous amorphous solid
water (p-ASW) \citep{Amiaud2006}. The smaller the fraction of molecules on a disordered surface, the higher
the binding energy. Except in a few cases,  in the ISM the abundance of molecules in water ices  is very
low \citep{Boogert2015}. Thus, in order to interpret observational data correctly and to predict abundances
using simulations, it is important to have the appropriate value of the binding energy of molecules on ices.
In this contribution, we report the careful measurement of binding energies of simple molecules (N$_2$, O$_2$,
CO, CH$_4$, and CO$_2$) on np-ASW at submonolayer (subML) and monolayer (ML) coverage, from $\sim$ 0.001 to 1 ML\@.
We also studied the desorption of N$_2$ and CO from p-ASW, and of NH$_3$ from crystalline water ice. We derived a simple formula to estimate the binding
energy change vs.\ submonolayer coverage; we then used this formula in  a gas-grain chemical network simulation
of a dense cloud and subsequent warm-up mimicking the hot-core/hot-corino phase.

The paper is structured as follows. In the next section we give a description of the apparatus and of the
experimental methods. In Section~\ref{sec:result}  we present the data and their analysis. From this analysis
we obtain a  simple function of the binding energy vs.\ coverage which, in Section~\ref{sec:astro}, is used
in a simulation of the chemical evolution of a dense cloud. In Section~\ref{sec:summary} we summarize the effect
of the new binding energy determinations on abundances of the evolution of a dense cloud.

\section{Experimental}
\label{sec:exp}
A detailed description of the apparatus can be found elsewhere \citep{Jing2013,He2015b,He2016a}, here we briefly summarize the main features that are most relevant for the measurement of the binding energy of molecules to water ice.
The experiments were carried out in a 10 inch diameter ultra-high vacuum chamber (``main chamber''). A pressure as low as $1.5\times10^{-10}$ torr is reached routinely after bake-out. During molecule deposition, the operating pressure is $\sim 3\times 10^{-10}$ torr. At this low pressure, background deposition is negligible. A 1~cm$^{2}$ gold coated copper disk substrate is located at the center of the chamber. It can be cooled down to 8~K by liquid helium or heated to 450~K using a cartridge heater. A Lakeshore 336 temperature controller with a calibrated silicon-diode (Lakeshore DT670) is used to measure and control the substrate temperature with an uncertainty of less than 0.5~K. A Hiden Analytic triple-pass quadrupole mass spectrometer (QMS) mounted on a rotary platform records the desorbed species from the sample or measures the  composition of the incoming beam. A Teflon cone is attached to the entrance of the QMS detector in order to maximize the collection of molecules desorbed or scattered from the surface. It has also the function of rejecting molecules  desorbing from other parts of the sample holder. The QMS is placed at $42\degree$ from the surface normal, while the incident angle of the molecular beam is $8\degree$ and on the opposite side---with respect to the surface normal---of the QMS detector. At the back of the sample, there is a gas capillary array for water vapor deposition. The capillary array is not directly facing the sample holder in order to obtain a deposition of water vapor from the background. Distilled water underwent at least three freeze-pump-thaw cycles to remove dissolved air. A leak valve is used to control the water vapor flow into the main chamber.
Three types of ices are used in this study, porous amorphous solid water (p-ASW), non-porous amorphous solid water (np-ASW), and crystalline water ice (CI).
During water vapor deposition, we regulated the vapor pressure to be $\sim 5\times 10^{-7}$ torr, which corresponds to a deposition rate of 0.5~ML/s. This rate is close to the slowest deposition rate used by \citet{Bossa2015}. The ice thickness is calculated by integration of the chamber pressure with time, assuming $1\times10^{-6}$ torr$\cdot$s exposure corresponds to 1 monolayer (ML). For experiments carried out in this study, the ice thickness is 100 ML\@. Porous amorphous ice is prepared by background deposition of water vapor when the substrate is at 10~K, followed by an annealing at 70~K for 30 minutes and by cooling back down to $\sim$15~K before the molecules are deposited on the surface using the molecular beam. Then in the TPD experiment  the temperature of the surface is raised linearly with time and the molecules coming from the surface are detected by the QMS\@.  In the  TPD experiments, the highest surface temperature for the p-ASW was kept below 70~K in order to maintain approximately the same surface morphology. Np-ASW  is prepared when the substrate temperature is at 130~K, followed by annealing at 130~K for 30 minutes. TPD experiments on np-ASW were performed at a temperature always below 130~K. CI was prepared by annealing a np-ASW sample at $\sim$145~K for 10 minutes. A Fourier Transform Infrared spectrometer (FTIR) in the RAIRS (Reflection Absorption InfraRed Spectroscopy) configuration is used to check the ice structure.
Connected to the main chamber are two highly collimated three-stage molecular beam lines. Gas flow into the beam was controlled by an Alicat MCS-5 mass flow controller.
In the third stage of the beam line there is a flag controlled by a stepper motor to accurately control the beam opening time. An exposure time as short as 0.5 second can be programmed with an uncertainty of 0.05 second. The list of TPD experiments are shown in Table~\ref{tab:list_exp}. The surface temperature at which the beam deposition takes place (deposition temperature) was chosen so that the sticking is unity \citep{He2016a}. The NH$_3$ TPD was performed on crystalline ice because NH$_3$ desorb at temperature where water ice becomes crystalline.
In the TPD the highest temperature was set to 145~K, at which the desorption of NH$_3$ is complete and water ice evaporation is still slow.

\begin{table}
\centering
\caption{List of temperature programmed desorption (TPD) experiments performed. The deposition dose ranges from below 1\% ML to above 1 ML\@. The temperature ramp rate during TPD is 0.5~K/s; np-ASW= non-porous amorphous solid water; p-ASW=porous amorphous solid water.}
\label{tab:list_exp}
\begin{tabular*}{0.45\textwidth}{@{\extracolsep{\fill} }ccc}
  \toprule
adsorbate & ice morphology & deposition temperature (K) \\
\midrule
CO        & p-ASW         & 15                         \\
CO        & np-ASW       & 15                         \\
N$_2$        & p-ASW         & 15                     \\
N$_2$        & np-ASW        & 15                     \\
O$_2$        & np-ASW       & 20                       \\
CH$_4$       & np-ASW        & 20                         \\
CO$_2$       & np-ASW       & 60                         \\
NH$_3$       & crystalline    & 50
\end{tabular*}
\end{table}

\section{Results and Discussion}
\label{sec:result}
\subsection{CO Desorption from Porous Amorphous Solid Water }
\begin{figure}
  \epsscale{1.1}
  \plotone{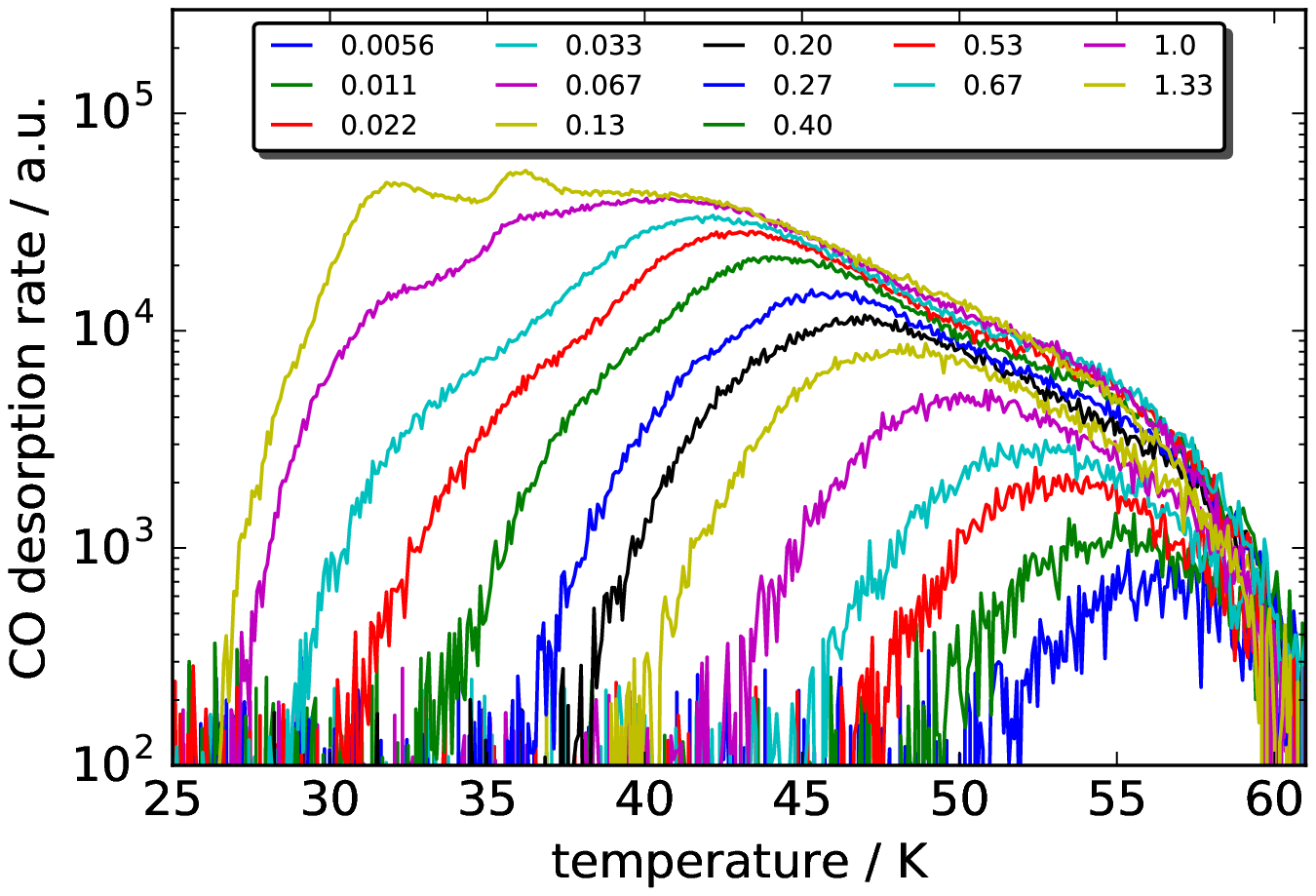}
  \caption{TPD spectra of CO from 100~ML of np-ASW\@. The temperature at which CO is deposited is 15~K. The TPD ramp rate is 0.5~K/s. Coverage in unit of ML is in the inset. }
\label{fig:CO_TPD_porous}
\end{figure}

The TPD spectra of CO from p-ASW are shown in Figure~\ref{fig:CO_TPD_porous}. Compared with TPD spectra in the literature, this work focuses on the submonolayer regime, especially on very low coverages. In the figure, CO coverages varies from 0.56\% ML to 1.33 ML\@.
The coverage is determined by identifying the transition from submonolayer to multilayer, similar to the method in \citet{Smith2016}. In the submonolayer regime, the TPD peak temperature decreases with coverage. At more than 1 ML, the TPD peak temperature increases with coverage, and shows zeroth order desorption with overlapping leading edges. A shape change is also seen at this transition. For experiments performed in this work, the absolute uncertainty of coverage determination is about 30\%. The coverage of other traces is obtained accordingly based on the deposition time. For p-ASW, it takes 30 minutes of CO exposure to fully cover the surface area, while for np-ASW, only 3 minutes of exposure are needed. This indicates that the surface area in a 100 ML p-ASW as prepared in our experimental condition is 10 times as that of np-ASW\@. During molecules deposition, the  beam intensity was stable and varies by less than 1\%. The sticking coefficient may also affect the amount of molecules on the surface. To eliminate the uncertainty induced by sticking, we set the surface temperature of deposition to be low enough so that the sticking coefficient during deposition is always unity~\citep{He2016a}.

\subsection{Direct Inversion Method}
The TPD traces in Figure~\ref{fig:CO_TPD_porous} share a common trailing edge, which indicates that the diffusion rate of CO is high and equilibrium diffusion state is reached during the desorption \citep{He2014a}. Molecules tend to occupy the deep adsorption sites (with higher binding energy) before they fill shallow sites (with lower binding energy). In this scenario, the direct inversion method---which is based on particles diffusing and occupying preferably deep adsorption sites---was used previously \citep[e.g.][]{He2014a,Smith2016} and works  well for extracting the coverage dependent binding energy distribution. A detailed discussion of the direct inversion method and the diffusion process on surface is available in \citet{He2014a}.  We start from the Polanyi-Wigner rate equation; assuming first order desorption, the desorption rate can be written as:
\begin{equation}
 \frac{\dif \theta(E_{\mathrm{D}},t)}{\dif t}= -\nu \theta
(E_{\mathrm{D}},t)\exp \left(-\frac{E_{\mathrm{D}}}{k_{\mathrm{B}}T(t)}\right)
\label{eq:p-w}
\end{equation}
where $\nu$ is the desorption pre-exponential factor that depends on the substrate and adsorbate; hereafter we use a widely accepted value of $10^{12}$ s$^{-1}$; $\theta (t)$ is the coverage defined as percentage of 1 ML, i.e., the number of adsorbate particles divided by the number of adsorption sites on the surface, $E_{\mathrm{D}}$ is the binding energy, $k_{\mathrm{B}}$ is Boltzmann constant, $T(t)$ is the temperature of the surface. The coverage dependent $E_{\rm D}(\theta)$ can be calculated for each TPD trace as follows:
\begin{equation}
 E_{\mathrm{D}}(\theta)=-k_{\mathrm{B}}T\ln\left(-\frac{1}{\nu \theta}\frac{\dif \theta}{\dif t}\right)
 \label{eq:invert}
\end{equation}

\begin{figure}
  \epsscale{1.1}
  \plotone{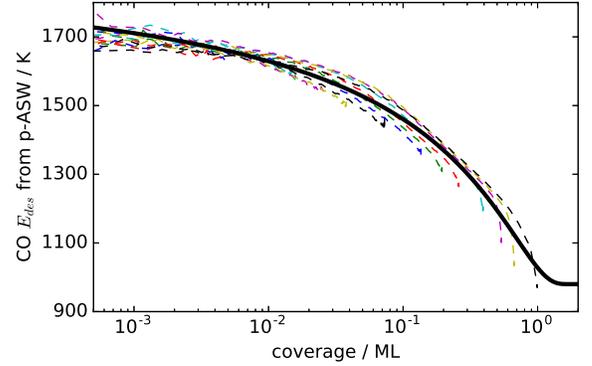}
  \caption{Surface coverage dependent binding energy distribution of CO on p-ASW obtained from a direct inversion method. The dashed lines are the inverted TPD traces in Figure~\ref{fig:CO_TPD_porous} using Eq~\ref{eq:invert}, the solid line is the fitting using formula in Eq~\ref{eq:fit} and parameters in Table~\ref{tab:fitting_para}.}
\label{fig:CO_porous_Edes}
\end{figure}

This inversion is applied to each TPD trace in Figure~\ref{fig:CO_TPD_porous}. The resulting $E_{\mathrm{D}}(\theta)$ distribution is shown in Figure~\ref{fig:CO_porous_Edes}.
It can be seen that the $E_{\mathrm{D}}(\theta)$ obtained from different TPD traces overlap well; this suggests that the direct inversion is a reliable method to obtain the binding energy distribution, giving that the trailing edge of the TPD traces overlap.
For easier incoporation of the $E_{\rm D}$ in gas-grain models, we used a function to fit the $E_{\mathrm{D}}(\theta)$ distribution:
\begin{equation}
 E_{\mathrm{D}}(\theta)= E_1 + E_2 \exp \left(-\frac{a}{\max(b-\lg (\theta), 0.001)}\right)
 \label{eq:fit}
\end{equation}
where $E_1$, $E_2$, $a$, and $b$ are fitting parameters.  $E_1$ is the binding energy for $\theta > 1$ ML, while $E_1+E_2$ is the binding energy when $\theta$ approaches zero. This function can fit the binding energy at both the submonolayer and multilayer coverage by forcing the multilayer $E_{\mathrm{D}}(\theta)$ value to be a constant $E_1$ using the $\max$ function. The calculated binding energy distributions with the fitting function in Equation~\ref{eq:fit} are shown in Figure~\ref{fig:CO_porous_Edes}. The fitting parameters are shown in Table~\ref{tab:fitting_para}.

\begin{figure}
  \epsscale{1.1}
  \plotone{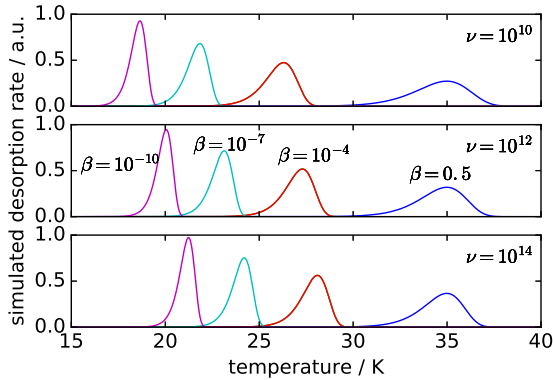}
  \caption{Effect of using different values of the prefactor $\nu$ in the analysis of experimental data  and in the gas-grain model. See text.}
\label{fig:changing_nu}
\end{figure}

So far we used $10^{12}$/s as the prefactor. However, in other works (i.e., \citet{Smith2016}), different values have been either assumed or obtained from fits. It is useful to analyze how different prefactors affect the simulation of interstellar chemistry. For simplicity, we consider an ideal case where the binding energy has a single value and the TPD peak is sharp.
Suppose in a TPD experiment with a heating ramp rate of 0.5~K/s, a desorption peak is at 35~K. From this one can calculate the binding energy $E_{\rm D}$ using the Polanyi-Wigner equation and assuming a certain prefactor $\nu$. These values of $E_{\rm D}$ and $\nu$ can then be used in models to predict at what temperature molecules  desorb from the interstellar grain surface. Since the heating up time scale of a dense cloud is in the order of $10^5$ years (see below), molecules desorb from the grain surface at a lower temperature than in the laboratory condition. In Figure~\ref{fig:changing_nu} we show the simulated desorption as a function of temperature using different prefactors and heating ramp rates. It can be seen that when the prefactor is changed by a factor of 100 (i.e., change from $10^{12}$ to $10^{14}$), even at the slowest ramp rate $10^{-10}$~K/s the difference in simulated TPD peak temperature is less than 1.4~K, which amounts to 7\% error. This error is usually smaller than the uncertainty in determining the binding energy. Therefore, the use of different prefactors is not significantly affecting the simulation of ISM environments, provided that the same prefactor is used  both in laboratory data analysis and in gas-grain models.

\begin{figure*}
\epsscale{1}
\plotone{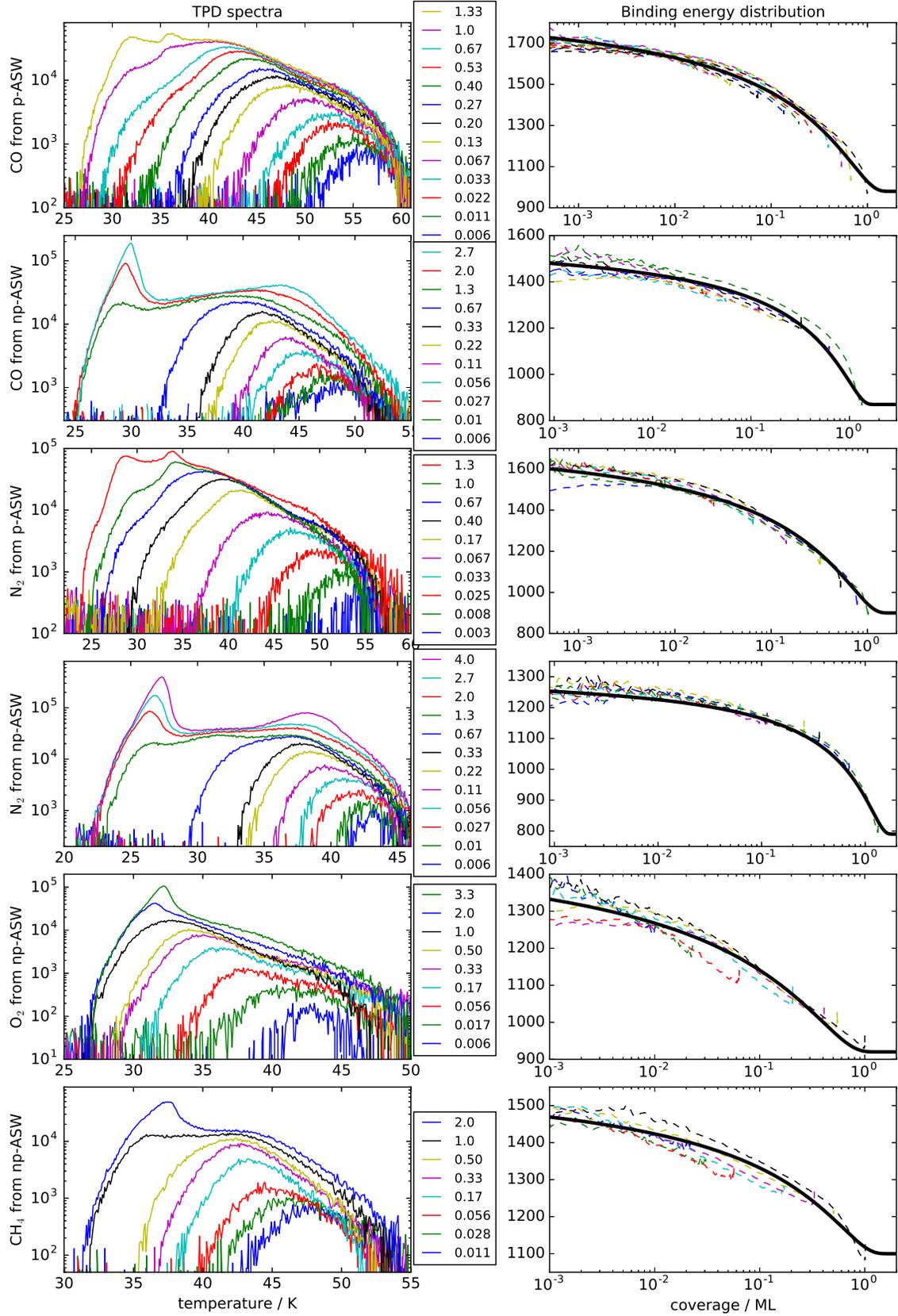}
  \caption{TPDs of CO and N$_2$ from  p-ASW, and of CO, N$_2$, O$_2$, CH$_4$, and D$_2$ from  np-ASW (left panel) with the corresponding binding energy distributions in K (right panel).  The coverage in ML is in the middle panel.}
\label{fig:6molecules}
\end{figure*}

The TPDs of N$_2$ from p-ASW, as well as N$_2$, CO, O$_2$, CH$_4$ from  np-ASW are shown in Figure~\ref{fig:6molecules}. We again apply the direct inversion method to each TPD trace from low coverage up to $\sim 1$ ML\@. Higher coverage curves are not included in the inversion result because the direct inversion method is not applicable for multilayer coverage. The binding energy distributions and the fittings are shown on the right panels. For TPDs of O$_2$ and CH$_4$ on np-ASW in the submonolayer regime, the trailing edges do not overlap well. Consequently, significant gaps are present in the inverted curves.

The coverage dependent binding energy distribution of H$_2$ on water ice is very important, since molecular hydrogen, as the most abundant molecules in the universe, plays an fundamental role in the physical and chemical evolution of interstellar clouds. However, it is very challenging to measure H$_2$ TPDs especially at low coverage because of the high background of H$_2$ even in a vacuum chamber under ultra-high vacuum conditions (P $\sim 10^{-10}$ torr). D$_2$ TPDs are typically measured instead. Here we use the D$_2$ TPDs on np-ASW by \citet{Amiaud2006}, as shown in Figure~\ref{fig:D2_TPD_compact}. We follow similar direct inversion procedures as above to obtain the binding energy distribution, as is shown in Figure~\ref{fig:D2_compact_Edes}. The same group recently performed TPDs of H$_2$, D$_2$, and HD on p-ASW, and compared the peak temperature differences between these three molecules \citep{Amiaud2015}. Based on Figure 2 in \citet{Amiaud2015}, the D$_2$ and H$_2$ desorption peaks are at 16 and 14~K, respectively. A calculation of the binding energy turns out that the ratio $E_{\rm D,H_2}/E_{\rm D,D_2} = 0.87$. We then assume that this ratio also applies to np-ASW and multiply the D$_2$ binding energy distribution by 0.87 to represent the binding energy distribution of H$_2$.

\begin{table}
\centering
\caption{Fitting parameters for the $E_{\rm D}(\theta)$ distribution as in Eq.~3. }
\label{tab:fitting_para}
\begin{tabular*}{0.42\textwidth}{@{\extracolsep{\fill} }ccccc}
  \toprule
    & E$_1$   & E$_2$  & a    & b   \\
    \midrule
CO (np-ASW)  & 870  & 730 & 0.6  & 0.3 \\
N$_2$ (np-ASW)  & 790  & 530 & 0.45 & 0.3 \\
O$_2$ (np-ASW)  & 920  & 600 & 1.2  & 0.2 \\
CH$_4$ (np-ASW) & 1100 & 500 & 1    & 0.3 \\
D$_2$ (np-ASW)  & 370  & 210 & 0.7  & 0.2 \\
H$_2$ (np-ASW)  & 322  & 183 & 0.7  & 0.2 \\
\midrule
CO (p-ASW)  & 980  & 960 & 0.9  & 0.3 \\
N$_2$ (p-ASW)  & 900  & 900 & 0.9 & 0.3 \\
\bottomrule
\end{tabular*}
\end{table}

\begin{figure}
  \epsscale{1}
  \plotone{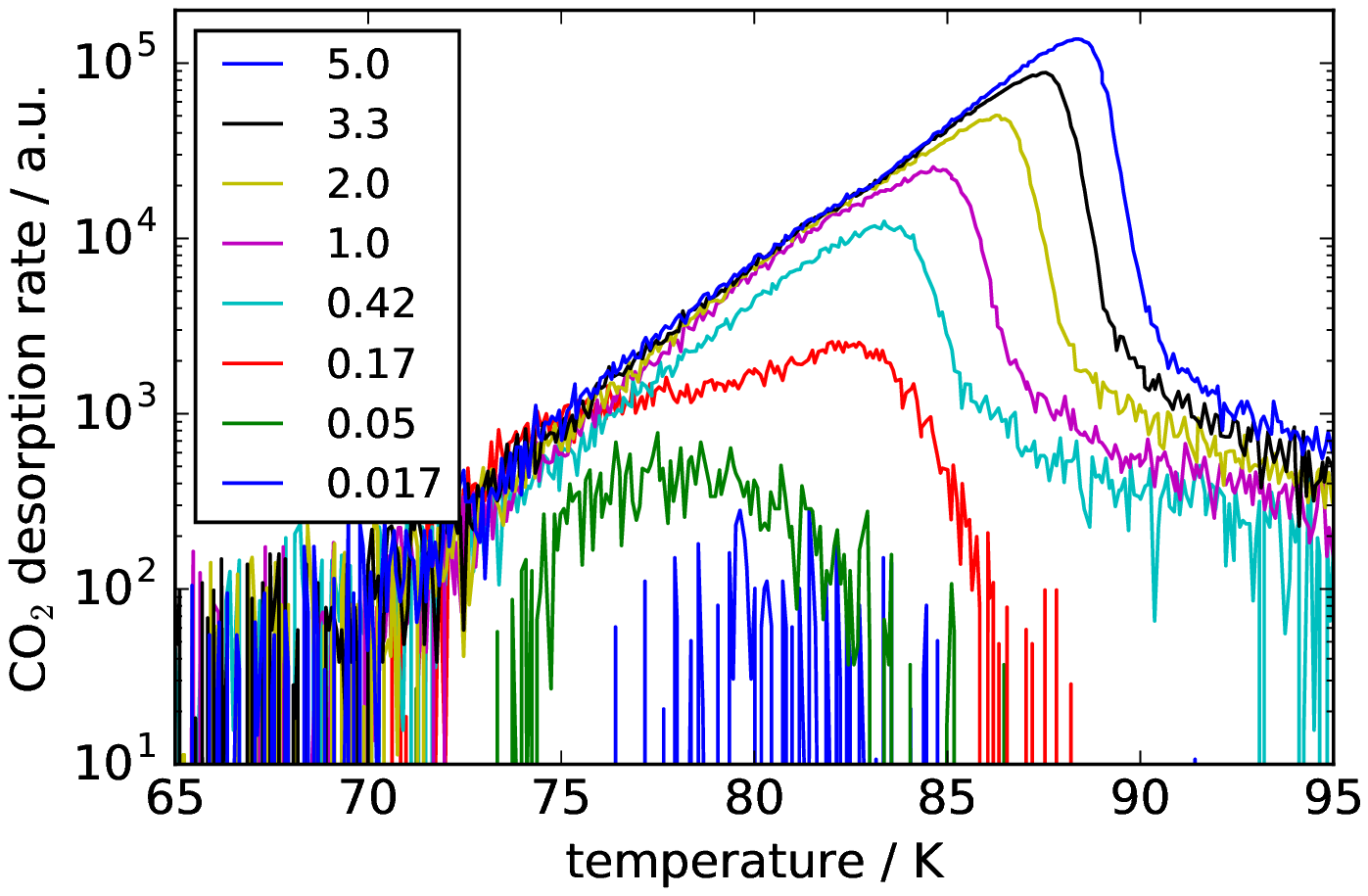}
  \caption{CO$_2$ TPDs from  np-ASW\@. The surface coverage is shown in the inset. CO$_2$ is deposited when the ice is at 60~K. During the TPD the heating ramp rate is 0.5~K/s.}
\label{fig:CO2_TPD_compact}
\end{figure}

Figure~\ref{fig:CO2_TPD_compact} shows the TPDs of CO$_2$ from np-ASW\@.  Zeroth order like traces are seen even at very low deposition length, indicating CO$_2$ forms clusters.  Surface coverage cannot be estimated by identifying the change in TPD shape. Instead, we assume that it takes the same amount CO$_2$ as CO (beam flux times deposition length) to cover 1 ML surface area. The beam flux is accurately controlled by the Alicat Mass flow controller and the correction factor for the specific gas is already taken into account by the controller. This is a fair assumption, considering that for all other species we measured, it takes 3 minutes of exposure to fully cover the first monolayer of np-ASW surface. The binding energy can be calculated using Polanyi-Wigner equation to be 2320~K, and it is coverage insensitive. The clustering behavior of CO$_2$ may help to explain the observed segregation of CO$_2$ in ices \citep{Oberg2009}. This issue is still open. \citet{Karssemeijer2014} used a new CO$_2$-H$_2$O potential to calculate the adsorption properties of CO$_2$ on amorphous solid water and crystalline ice. They found no evidence of clustering in contradiction with experiments by \citet{Edridge2013} of CO$_2$ adsorption on ASW and graphite. The segregation of molecules in water ice will be the focus of a forthcoming paper.

\begin{figure}
  \epsscale{1.1}
  \plotone{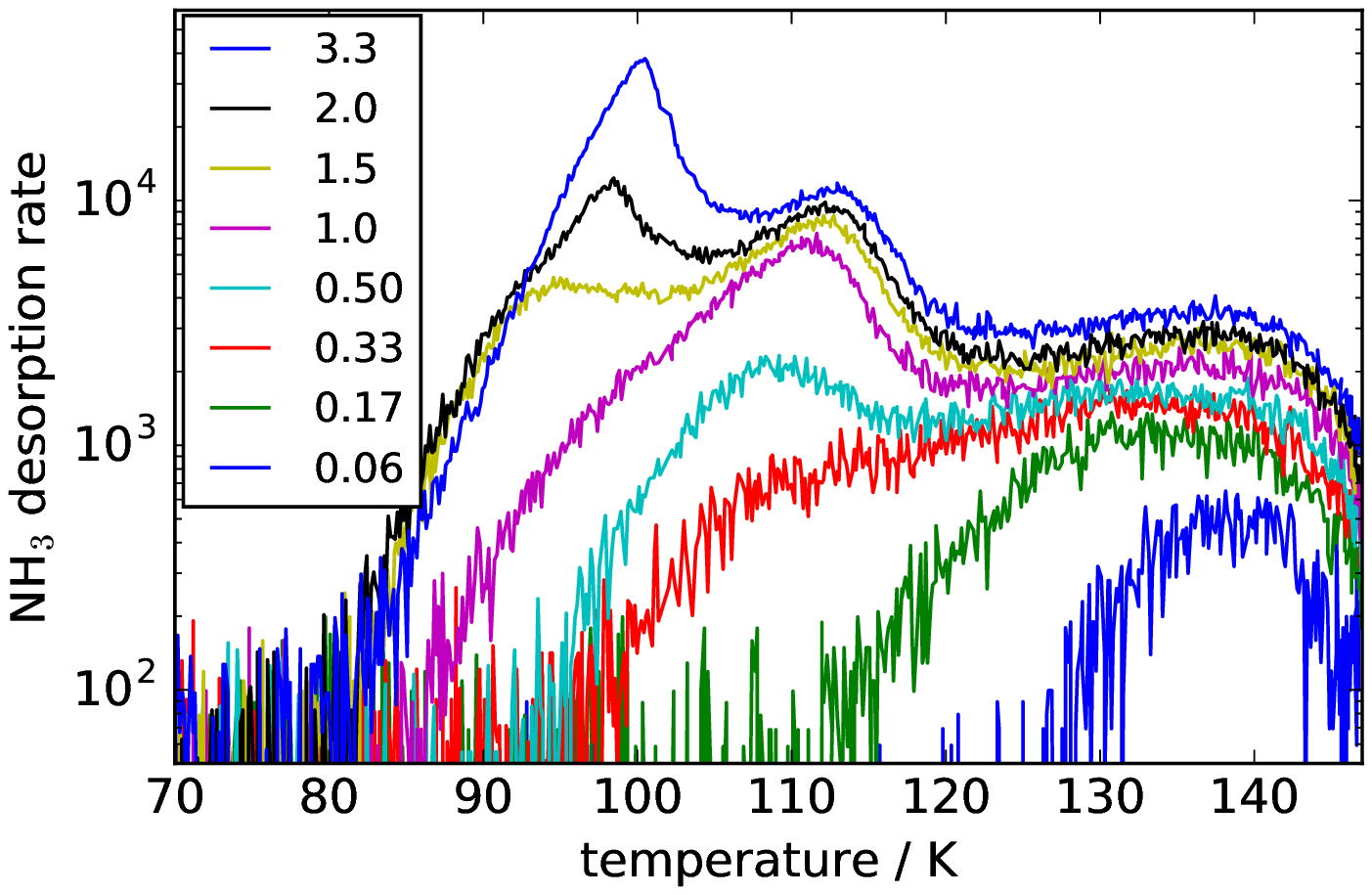}
  \caption{NH$_3$ TPDs from crystalline water ice (CI). The surface coverage is shown in the inset. NH$_3$ is deposited when the ice is at 50~K. During the TPD the heating ramp rate is 0.5~K/s. The ice is heated only to 145~K because at a higher temperature water begins to desorb significantly. }
\label{fig:NH3_TPD_crys}
\end{figure}

\begin{figure}
  \epsscale{1.1}
  \plotone{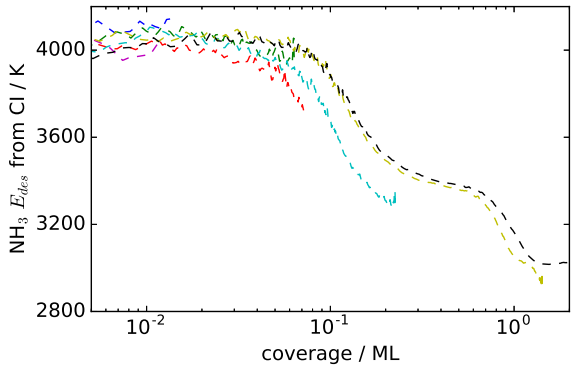}
  \caption{Surface coverage dependent binding energy distribution of NH$_3$ on crystalline water ice (CI) obtained from a direct inversion of TPD traces in Figure~\ref{fig:NH3_TPD_crys}.}
\label{fig:NH3_crys_Edes}
\end{figure}

\begin{figure}
  \epsscale{1.1}
  \plotone{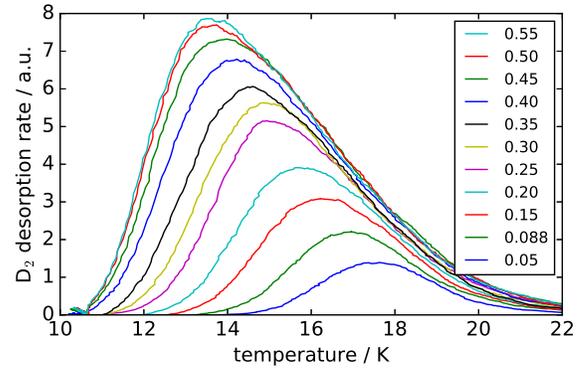}
  \caption{D$_2$ TPDs from np-ASW\@. The surface coverage is shown in the inset. Data taken from \citet{Amiaud2006}. }
\label{fig:D2_TPD_compact}
\end{figure}

\begin{figure}
  \epsscale{1}
  \plotone{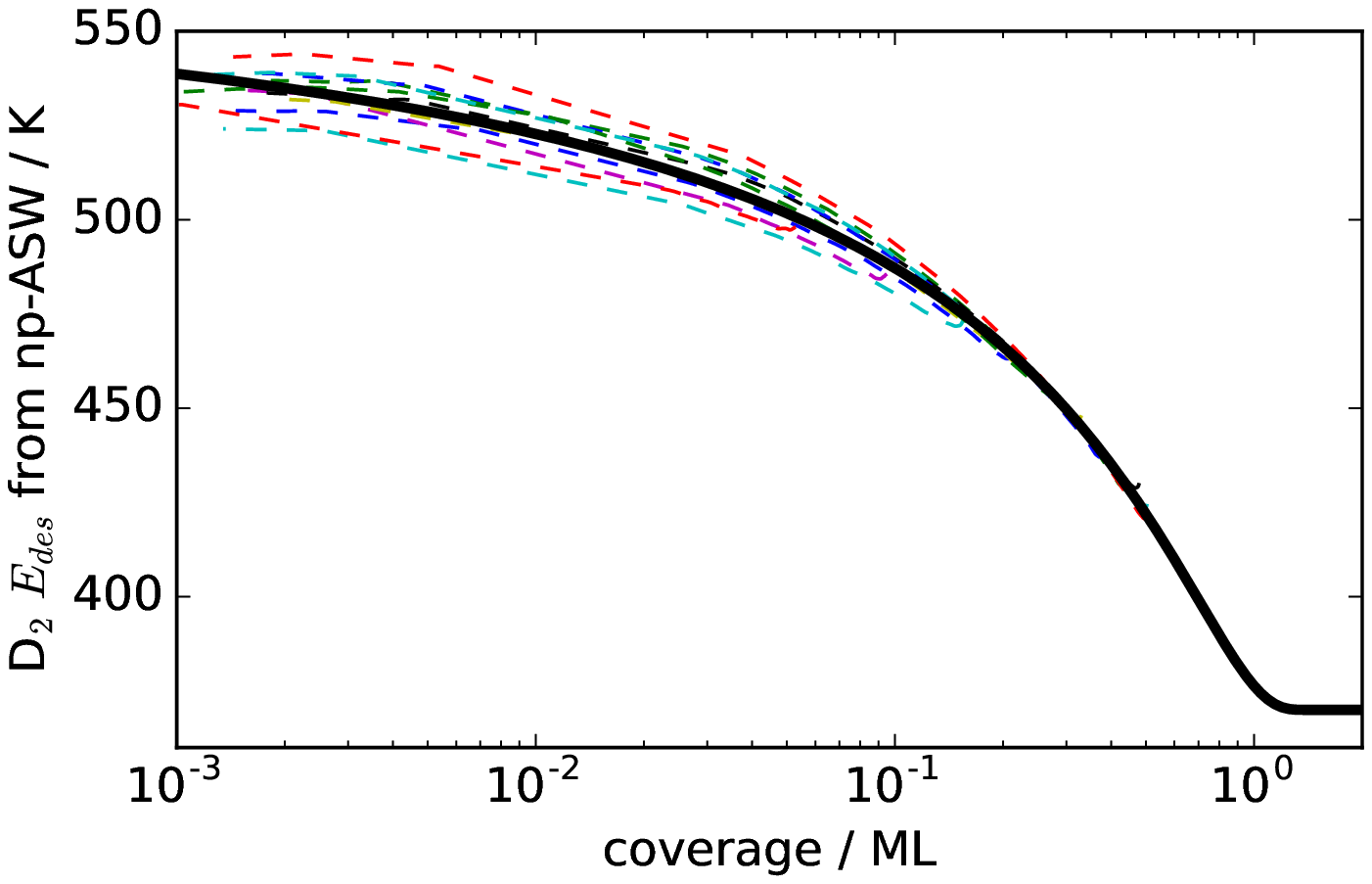}
  \caption{Surface coverage dependent binding energy distribution of D$_2$ on np-ASW obtained from a direct inversion of TPD traces in Figure~\ref{fig:D2_TPD_compact}. Data taken from \citet{Amiaud2006}. }
\label{fig:D2_compact_Edes}
\end{figure}

Figure~\ref{fig:NH3_TPD_crys} shows the TPDs of NH$_3$ from CI\@. Three distinct peaks are shown in the TPDs, at $\sim 140$~K, $\sim 110$~K, and $\sim 100$~K respectively. The 100~K peak is due to zeroth order desorption at multilayer coverage. The origin of the other two peaks is unknown and its explanation is out of scope of this work. A direct inversion is applied to the TPD traces and the so obtained binding energy distribution is shown in Figure~\ref{fig:NH3_crys_Edes}. This distribution cannot be fitted with the expression in Eq~\ref{eq:fit}. When the coverage of NH$_3$ is low, ammonia binds strongly to water ice and desorbs at a temperature slightly lower than the water desorption temperature. A more relevant measurement would be the desorption of NH$_3$ from the surface of a bare silicate or a carbonaceous material, but this is outside the focus of this work.

\section{Astrophysical Implications}
\label{sec:astro}
Grain surface chemistry plays a crucial role in the formation of a wide range of molecules including the simplest and most abundant molecule, H$_2$. Once a gas species adsorbs on  the grain surface it will hop on the surface due its thermal energy. Reactions occur via Langmuir-Hinshelwood (L-H), Eley-Rideal (E-L), or hot-atom (H-A) mechanism \citep{Vidali2013}. Here we only consider the L-H mechanism, i.e., reactants thermally migrate around the grain surface over the diffusion energy barrier $E_{\rm d}$ between sites until they meet at a binding site. Migration may also occur due to quantum mechanical tunneling. Initially it was postulated that mobility of  light atoms on a grain surface at low temperature was due to quantum tunneling, but laboratory experiments suggest that the thermal hopping plays a bigger role \citep{Pirronello1997b,Vidali2013,Hama2013}. The diffusion energy barriers define the rates at which reactions take place.

The rate of hopping at a given temperature $T$ is given by:
\begin{equation}
A_i= \nu_i\exp(-E_{\rm D}(i)/k_{\rm b}T),
\end{equation}
where, $k_{\rm b}$ is the Boltzmann constant and $\nu_i$ is the typical vibrational frequency, given by,
\begin{equation}
\nu_i= \sqrt{\frac{2 g_{\rm s} E_{\rm D}(i)}{\pi^2 m_i}},
\end{equation}
where, $g_{\rm s}$ is surface density of sites on a grain, $m_i$ is the mass of the $i$-th species and $E_{\rm D}(i)$
is the binding energy for desorption. The typical values of vibrational energies are in the range of
$10^{12}$--$10^{13}$ s$^{-1}$. It should be noted that in the direct inversion of the TPD traces, the $\nu$ value was taken to be $10^{12}$ s$^{-1}$ for all species. The small difference in $\nu$ values does not affect the result significantly. If one reactant finds another during hopping, they will recombine and form a
molecule. However sometimes adsorbed species may desorb back into the gas phase without reacting due
to variety of desorption mechanisms. The thermal desorption rate is given by:
\begin{equation}
W_i= \nu_i\exp(-E_{\rm D}(i)/k_{\rm b}T).
\end{equation}
Thus, $E_{\rm D}(i)$ and $E_{\rm d}(i)$ are two very crucial parameters that control grain surface chemistry once a gas phase species is adsorbed
on the surface.

Present day astrochemical model assumes a fixed $E_{\rm D}$ and $E_{\rm d}$ for the study of formation of grain
surface species. From the experiments discussed in the previous sections, it is clear
that these parameters are function of coverage. In the submonolayer regime the binding
energy is significantly higher than in the monolayer regime; the consequence is  that
while molecules are kept on the grain surface longer, the formation rate is reduced due to
slower hopping. We employed a gas-grain simulation to examine the impact of the experimental results that are
discussed here. The details of the simulation and chemical networks used are given in \citep{He2016a,Acharyya2015a,Acharyya2015b}.
\begin{table}
\centering
\caption{Model Names and Parameters }
\label{Table_Model}
\begin{tabular*}{0.45\textwidth}{@{\extracolsep{\fill} }ccccc}
\toprule
Name  & T$_{\rm Grain}$ & Heating Rate            & Density             & Binding \\
      &  (K)        & (K/year)                & cm$^{-3}$           & Energy  \\
\midrule
C1N   & 10          & No                      & 2 $\times$ $10^{4}$ & Eq. 3   \\
C1O   & 10          & No                      & 2 $\times$ $10^{4}$ & Old     \\
C2N   & 10          & No                      & 1 $\times$ $10^{5}$ & Eq. 3   \\
C2O   & 10          & No                      & 1 $\times$ $10^{5}$ & Old     \\
\midrule
W1FN  & 10--200    & 190/5 $\times$ $10^{4}$ & 2 $\times$ $10^{4}$ & Eq. 3   \\
W1FO  & 10--200    & 190/5 $\times$ $10^{4}$ & 2 $\times$ $10^{4}$ & Old     \\
W2FN  & 10--200    & 190/5 $\times$ $10^{4}$ & 1 $\times$ $10^{5}$ & Eq. 3   \\
W2FO  & 10--200    & 190/5 $\times$ $10^{4}$ & 1 $\times$ $10^{5}$ & Old     \\
\midrule
W1SN  & 10--200    & 190/1 $\times$ $10^{6}$ & 2 $\times$ $10^{4}$  & Eq. 3  \\
W1SO  & 10--200    & 190/1 $\times$ $10^{6}$ & 2 $\times$ $10^{4}$  & Old    \\
W2SN  & 10--200    & 190/1 $\times$ $10^{6}$ & 1 $\times$ $10^{5}$  & Eq. 3  \\
W2SO  & 10--200    & 190/1 $\times$ $10^{6}$ & 1 $\times$ $10^{5}$  & Old    \\
\bottomrule
\end{tabular*}
\end{table}

\subsection{Model Parameters}
We ran two classes of models: in one we considered the old fixed binding energy for CO, N$_2$, O$_2$,
CH$_4$, H$_2$ and atomic oxygen and in another we have used Equation~\ref{eq:fit} for CO, N$_2$, O$_2$,
CH$_4$, H$_2$ and  1850~K  for the atomic oxygen binding energy \citep{He2015b}. The sticking was treated the same way as that in \citet{He2016a}. We ran six models for each
class by varying the density and gas and grain temperature. Two models for each class were run by keeping
the gas and grain temperature fixed at 10~K. They differ by the initial density; in one we considered
the hydrogen number density ($n_{\rm H}$) $2\times 10^4$ cm$^{-3}$ and in other $10^5$ cm$^{-3}$. We
designate these two models as C1O and C2O (``O'' for old energies) and C1N and C2N (``N'' for new energies and
Equation~\ref{eq:fit}). Then we ran another four models (``W'' for warm-up) for each class in which we
warmed up the grain and gas and increased the temperature from 10~K to 200~K at a linear rate. We have
considered two different densities, $n_{\rm H}=2\times 10^4$ (W1) and $10^5$ cm$^{-3}$ (W2) and two
different heating rates for increasing the gas and grain temperature from 10~K to 200~K:\@ fast heating
(``F'') in $5\times 10^4$ years  and slow heating (``S'') in $10^6$ years. Thus we have W1FO and W2FO
(fast), W1SO and W2SO (slow) models with old binding energy and W1FN and W2FN (fast), W1SN and W2SN (slow)
models for which Equation~\ref{eq:fit} is used.

\begin{figure}
  \epsscale{1}
  \plotone{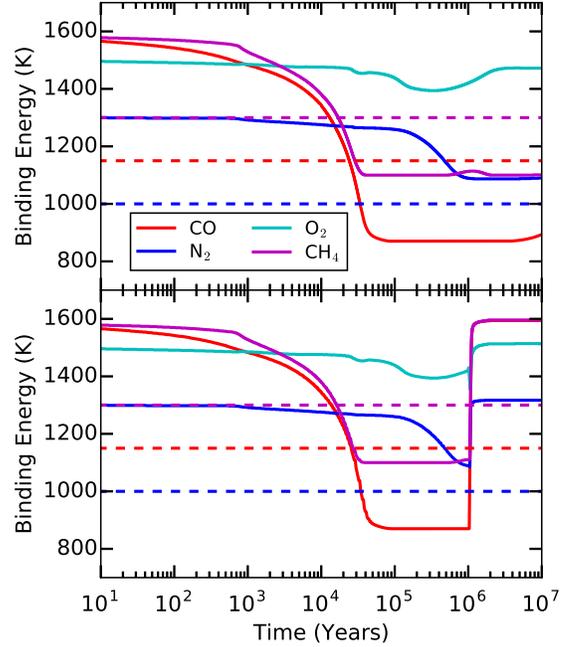}
\caption{Time variation of binding energy as calculated during the model run for: C1N (solid lines)
and C1O (dashed lines) model (top panel);   W1FN (solid lines) and W1FO model (dashed lines) (bottom panel).}
\label{fig-astro1}
\end{figure}

\subsection{Results}
In Figure~\ref{fig-astro1} (top), solid curves show the variation of binding energy as a function of time during the
calculation for the C1N model and dashed lines show fixed binding energy used for the C1O model. Since both
O$_2$ and N$_2$ have the same fixed binding energy (1000~K) in the C1O model only N$_2$ is shown.
For CO, the binding energy for the C1N model (solid red line) is high up to $3\times 10^4$ years, and after that
it is lower compared to the fixed binding energy of 1150~K (red dashed line). For O$_2$ (solid cyan line)
and N$_2$ (solid blue line), for all times the binding energy in the C1N model is higher than in the C1O model (dashed
blue line) which essentially mean that these two species always desorb in the sub-monolayer regime. The time
variation of the CH$_4$ binding energy  shows that for  time greater than $2\times 10^4$ years,
the binding energy (solid magenta line) in model C1N is lower than in  C1O (dashed magenta line). Figure~\ref{fig-astro1} (bottom),
shows the binding energy variation in the W1FN (solid curves) model. Up to $10^6$ years, the binding energy profile of all the
species is the same as shown in Figure~\ref{fig-astro1} (top). However, once grains begin to warm up and the  surface
abundance decreases below the  monolayer regime, we can see that the binding energy starts to increase and attains the
maximum level that is sum of E$_1$ and E$_2$.

\subsubsection{CO}
\begin{figure}
  \epsscale{1.1}
  \plotone{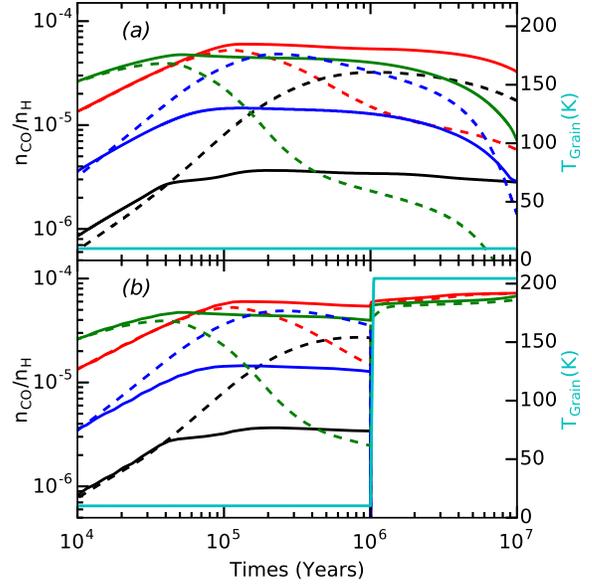}
\caption{The time variation of CO abundance is shown for different models. Solid and dashed lines are for
models with Eq.~3 and with the old binding energies, respectively. Red (gas) and black (grain) lines are
for $n_{\rm H}=2\times 10^4$ cm$^{-3}$ and green (gas) and blue (grain) lines  for $n_{\rm H}=1\times
10^5$ cm$^{-3}$. (a) is for cloud with fixed gas and grain temperature of 10~K, and,
(b) is warm up model in which grain temperature is raised from 10 to 200~K in $5\times 10^4$ years.
}
\label{fig-astro2}
\end{figure}

Figure~\ref{fig-astro2} shows the time variation of CO abundance for various models.
In the top panel the CO abundance for models C1O, C2O, C1N, and C2N is shown. For these models the
gas and grain temperatures are kept constant at 10~K (cyan dashed line).
In Figure~\ref{fig-astro2}a, the red curve shows the gas-phase CO abundance. When  Equation~\ref{eq:fit} is used for $n_{\rm H}$ =
$2\times 10^4$ cm$^{-3}$, the CO gas-phase abundance is significantly higher than in the model with the old and fixed value of the energy (dashed
red curve). This is mainly due to the fact that in the astronomical literature the binding energy for CO
is taken as 1150~K (shown in red dashed line in Figure~\ref{fig-astro1}a) whereas when Equation~\ref{eq:fit} is used
the CO binding energy remains close to 840~K for a time greater than $4\times 10^4$ years, see
 Figure~\ref{fig-astro1} (red curve). Thus, most of the time the CO binding energy is lower than
the one used in the literature. A lower binding energy of CO makes cosmic ray induced
desorption  rate \citep{Hasegawa1993} to be greater than the accretion rate of CO\@. The net effect is very
 little CO depletion. Consequently we can see that grain surface CO abundance in the  C1N model (solid black
line) is lower than in the CIO model (dashed black line). To investigate this matter further we ran a
model with $n_{\rm H}$ = $10^5$ cm$^{-3}$. The solid green line shows the CO gas-phase abundance; after $10^6$ years it
starts to go down much more rapidly than in the previous model but is still significantly lower than in the
C2O model (dashed green line). The solid blue line shows the grain surface CO abundance for $n_{\rm H} =
10^5$ cm$^{-3}$; it is still much lower compared to the one in the C2O model  (dashed blue line).

Figure~\ref{fig-astro2}b shows the time variation of the CO abundance for the warm up model with the fast heating rate.
For these models, the  gas and grain temperature is kept constant at 10~K up to $10^6$ years and then  is linearly
increased to 200~K in $5\times 10^4$ years. The solid red line shows the CO gas-phase abundance  for the W1FN model;
it is clear that with the W1FO
model there is hardly any change from the warm-up phase to the post warm-up phase.   Once the warm-up process begins and the grain temperature goes above 20~K,  thermal desorption
takes over and becomes the most dominant desorption process; by 30~K, grain surface CO is completely
evaporated back into gas phase. A similar trend is seen for surface CO\@.
In Figure~\ref{fig-astro2}b the solid red curve (dashed red line) shows the binding energy variation of CO with time for this model (old model with fixed 1150~K binding energy). It is clear that just above $10^6$ years CO attains
maximum binding energy of 1600~K in warm-up models since there is very little CO on the grain surface.

\begin{figure}
  \epsscale{1.1}
  \plotone{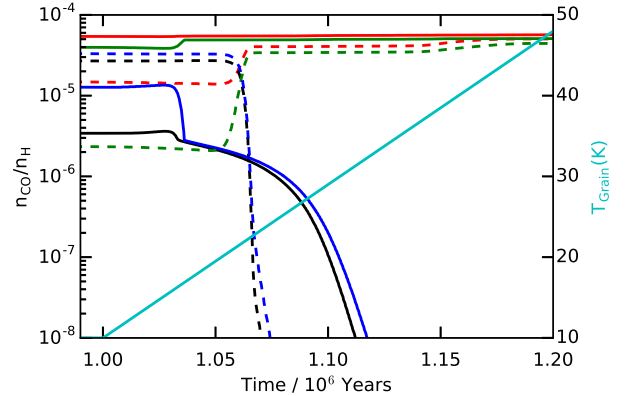}
\caption{Same as Figure~\ref{fig-astro1}b but the grain temperature is heated to 200~K in $10^6$ years; the early warm-up phase is enlarged to show the sub-monolayer regime desorption of CO.}
\label{fig-astro3}
\end{figure}

To understand CO desorption from the sub monolayer regime during warm up we ran models with a slow heating
rate; the time variation of CO abundance is shown in the Figure~\ref{fig-astro3}. Since the pre-collapse
behavior is the same as shown in the Figure~\ref{fig-astro2}, in this plot we only highlighted the portion of the
curve in which CO is desorbing due to the warm-up. It can be seen that the gas phase abundance of CO i.e., solid
red line ($n_{\rm H}=2\times 10^4$ cm$^{-3}$) and solid green line ($n_{\rm H}=1 \times 10^5$ cm$^{-3}$)
for the W1SN and W2SN models, starts to increase much earlier than in the W1SO and W2SO models (dashed red and green
line) due to the lower binding energy in the monolayer regime compared to the fixed binding energy. But once it
desorbed back to the gas-phase, the abundance profile is very similar. It is clear that grain surface CO
abundance for both the W1SN and W2SN models starts to decrease earlier than in W1SO and W2SO models but for
W1SN and W2SN models the decrease is much slower and we see a knee like decrease. This means that a small fraction
of CO is retained up to a much longer time and up to higher temperature ($\sim$40~K) when the binding energy
as a function of coverage is used. However, when heating rate is fast the effect is small
(Figure~\ref{fig-astro2}b).

\subsubsection{\texorpdfstring{N$_2$}{N2}}
\begin{figure}
  \epsscale{1.1}
  \plotone{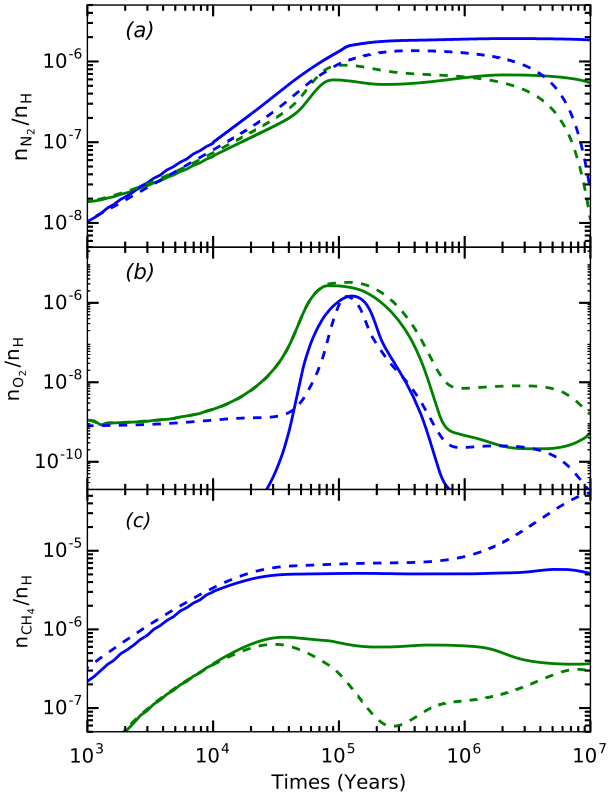}
\caption{Similar to Figure~\ref{fig-astro2}a but for N$_2$, O$_2$ and CH$_4$ and for $n_{\rm H} = 10^5$ cm$^{-3}$.}
\label{fig-astro4}
\end{figure}

Figure~\ref{fig-astro4}a shows the time variation of N$_2$ for the C2N and C2O models. Gas phase N$_2$ abundance
for the C2N model (solid green line) remains almost unchanged after $10^5$ years, similarly to the CO abundance profile,
whereas for the C2O model, the abundance goes down initially slowly and then sharply after $10^5$ years. If we now
look at  Figure~\ref{fig-astro1} and compare the binding energy used during the model run, it is clear that
for all times N$_2$ has a higher binding energy in the C2N model (solid blue line) than in the C2O model (dashed
blue line). Then the question  is why  gas-phase N$_2$ is more strongly enhanced in the new model. Looking at the gas phase
formation pathways of N$_2$ we found that at late times the most dominant pathway for N$_2$ formation in the
C2N model is via CO + N$_2$H$^+$ where as in the C2O model the most dominant pathway is N$_2$H$^+$ + e. Thus,
a significant high abundance of CO in the C2N model causes an enhanced abundance of N$_2$.
Surface N$_2$ in  the C2N model is also almost constant due to steady accretion from the gas phase.
Figure~\ref{fig-astro4} shows that the gas phase N$_2$ abundance for W2SN (solid green line) and W2SO
(dashed green line) models is very similar. However, for surface N$_2$, the desorption starts little later and takes longer
time in model W2SN (solid blue line) due to higher binding energy compared to the W2SO model (dashed blue line).

\begin{figure}
  \epsscale{1.1}
  \plotone{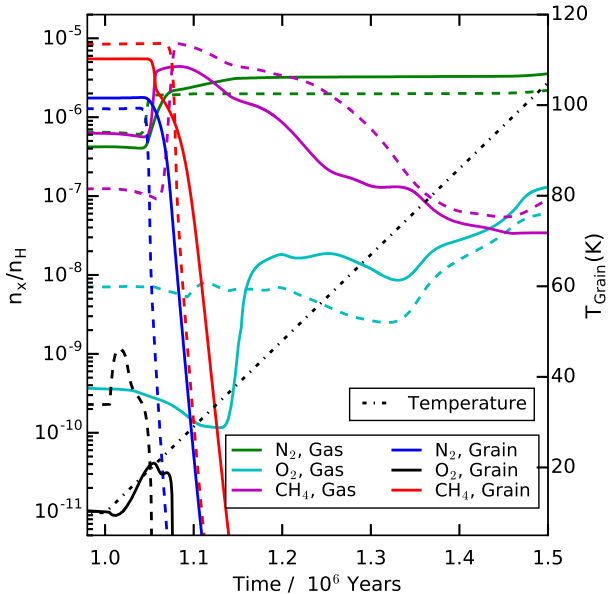}
\caption{Similar to Figure~\ref{fig-astro3} but for N$_2$, O$_2$ and CH$_4$.}
\label{fig-astro5}
\end{figure}

\subsubsection{\texorpdfstring{O$_2$}{O2}}
Figure~\ref{fig-astro4}b shows the  time variation of O$_2$ for the C2N and C2O models. Up to $10^5$ years
in both the C2N and C2O models the gas-phase abundance for O$_2$ is almost the same. When time is further increased
the abundance of O$_2$ in the C2N model starts to decrease compared to the C2O model, and above $10^6$ years nearly an
order of magnitude difference in abundance could be seen between these two models. It is clear from
Figure~\ref{fig-astro1} that the coverage dependent O$_2$ binding energy is significantly higher compared to
the fixed binding energy of 1000~K. Therefore we see more O$_2$ in the C2O model compared to the C2N model due to
higher non-thermal desorption rate owing to a lower binding energy.
Surface O$_2$ abundance  before $2\times 10^4$ years for the C2O model is significantly  higher
than in the C2N model. During this time the major source of O$_2$ on the grain surface is via O+O reaction. Due to the
very high binding energy of oxygen in  model C2N, there is no O$_2$ formation via this route. After $10^5$
years the major source of O$_2$ on the grain surface is via accretion from the gas phase. Thus during this
time the O$_2$ surface abundance in C2N and C2O models  is very similar. Once again we see a difference
at time greater than $10^6$ years. During this time the C2O model surface O$_2$ abundance is  due to the relatively
higher accretion rate caused by the higher gas-phase abundance of O$_2$ in this model.
We can see the same effect for W2SO and W2SN models in the Figure~\ref{fig-astro5}.

\subsubsection{\texorpdfstring{CH$_4$}{CH4}}
Figure~\ref{fig-astro4}c shows time variation of CH$_4$ abundance for C2N (solid green line) and C2O
(dashed green line) models. Up to $10^4$ years the gas-phase abundance remains very similar in both models.
After that, although the C2N model abundance remains almost constant,  C2O model abundance starts to decrease
up to $10^5$ years and then begins to increase again after 3 $\times 10^5$ years. At very late time the abundance
for both models is similar. The main reasons for this deviation is two-fold. First for the C2N model, the
binding energy used during the model calculation is lower than the C2O model between $10^4$ and $10^6$ years.
Therefore, net depletion in the  C2N model is much lower compared to the C2O model. And the second reason is
that the most dominant gas-phase formation pathways for CH$_4$ is via reaction with CO and CH$_5^+$. Thus more
CO in the C2N model will cause more CH$_4$ formation in this model. Similarly, due to the higher binding of
CH$_4$ in the C2O model there will be less non thermal desorption in the C2O model; this will cause an
increase in surface CH$_4$ abundance  at late times as evident in  Figure~\ref{fig-astro4}c (dashed blue line).
Figure~\ref{fig-astro5} shows the CH$_4$ abundance for the W2SN and W2SO models. We can see clearly that in the
W2SN model, as temperature is increased CH$_4$ starts to desorb earlier than in the W2SO model due
to its lower binding energy in the monolayer regime. On the other hand, although surface CH$_4$ in the W2SN
model also starts to decrease due to thermal desorption before the W2SO model, owing to its higher binding
energy in the sub-monolayer regime, it slows down and  takes longer time to complete the desorption.

\section{Summary}
In this work we measured the binding energy distribution of N$_2$, CO, O$_2$, CH$_4$, and CO$_2$  on non-porous amorphous solid water (np-ASW),
of N$_2$ and CO on porous amorphous solid water (p-ASW), and of NH$_3$ on crystalline water ice down to a fraction of 1\% of the layer. We found that CO$_2$ forms clusters on np-ASW surface even at very low coverage. This may help to explain the observed CO$_2$ segregation in ices. The binding energy of N$_2$, CO, O$_2$, and CH$_4$ decreases with coverage. The energy values in the low coverage limit are much higher than those commonly used in gas-grain astrochemical models. We found a simple empirical formula that gives the binding energy of a molecule as a function of the coverage. We then used this formula in a simulation of the time evolution of a dense cloud followed by a warm-up
as appropriate in a hot core or hot corino.  We found that for O$_2$ and N$_2$, desorption takes place from
the sub-monolayer regime for all models we considered, therefore effective desortion and hopping energies are higher compared to the
single energy values used in the current day astronomical models. For CO and CH$_4$ and for cold cloud models,
initially the effective binding energy remains significantly higher; then it gradually decreases to the value for  a monolayer
 as coverage is increased and remains at this value till to the end of the simulation. In the warm-up model, the binding energy increases with  grain temperature  and attains the maximum value
of $E_1+E_2$, see Table~\ref{tab:fitting_para}. Another important outcome is that during the slow warm-up, although the
desorption process starts earlier for species like CO and CH$_4$ due to lower value of binding energy at
monolayer regime, it takes longer time to complete compared to the single value binding energy due the
significantly higher sub-monolayer binding energy. Thus, a fraction of all these ices stays much longer on the
grain surface compared to the case of using a single value of the binding energy  as is currently done  in astrochemical models.
\label{sec:summary}
\section{Acknowledgments}
We would like to thank S M Emtiaz and Xixin Liang for technical assistance. This work was supported in part by
a grant to GV from NSF---Astronomy \& Astrophysics Division (\#1311958). K. A. would like to thank the support of local funds
from Physical Research Laboratory.

\bibliography{TPD}

\begin{thebibliography}{33}
\expandafter\ifx\csname natexlab\endcsname\relax\def\natexlab#1{#1}\fi

\bibitem[{{Acharyya} \& {Herbst}(2015)}]{Acharyya2015b}
{Acharyya}, K., \& {Herbst}, E. 2015, \apj, 812, 142

\bibitem[{{Acharyya} {et~al.}(2015){Acharyya}, {Herbst}, {Caravan}, {Shannon},
  {Blitz}, \& {Heard}}]{Acharyya2015a}
{Acharyya}, K., {Herbst}, E., {Caravan}, R.~L., {et~al.} 2015, Molecular
  Physics, 113, 2243

\bibitem[{{Amiaud} {et~al.}(2006){Amiaud}, {Fillion}, {Baouche}, {Dulieu},
  {Momeni}, \& {Lemaire}}]{Amiaud2006}
{Amiaud}, L., {Fillion}, J.~H., {Baouche}, S., {et~al.} 2006, \jcp, 124, 094702

\bibitem[{{Amiaud} {et~al.}(2015){Amiaud}, {Fillion}, {Dulieu}, {Momeni}, \&
  {Lemaire}}]{Amiaud2015}
{Amiaud}, L., {Fillion}, J.-H., {Dulieu}, F., {Momeni}, A., \& {Lemaire}, J.-L.
  2015, Physical Chemistry Chemical Physics (Incorporating Faraday
  Transactions), 17, 30148

\bibitem[{{Antczak} \& {Ehrlich}(2005)}]{Antczak2005}
{Antczak}, G., \& {Ehrlich}, G. 2005, Surface Science, 589, 52

\bibitem[{{Biham} {et~al.}(2001){Biham}, {Furman}, {Pirronello}, \&
  {Vidali}}]{Biham2001}
{Biham}, O., {Furman}, I., {Pirronello}, V., \& {Vidali}, G. 2001, \apj, 553,
  595

\bibitem[{{Bisschop} {et~al.}(2007){Bisschop}, {J{\o}rgensen}, {van Dishoeck},
  \& {de Wachter}}]{Bisschop2007}
{Bisschop}, S.~E., {J{\o}rgensen}, J.~K., {van Dishoeck}, E.~F., \& {de
  Wachter}, E.~B.~M. 2007, \aap, 465, 913

\bibitem[{{Boogert} {et~al.}(2015){Boogert}, {Gerakines}, \&
  {Whittet}}]{Boogert2015}
{Boogert}, A.~C.~A., {Gerakines}, P.~A., \& {Whittet}, D.~C.~B. 2015, \araa,
  53, 541

\bibitem[{{Bossa} {et~al.}(2015){Bossa}, {Mat{\'e}}, {Fransen}, {Cazaux},
  {Pilling}, {Robson Monteiro Rocha}, {Ortigoso}, \& {Linnartz}}]{Bossa2015}
{Bossa}, J.-B., {Mat{\'e}}, B., {Fransen}, C., {et~al.} 2015, \apj, 814, 47

\bibitem[{{Bruch} {et~al.}(2007){Bruch}, {Diehl}, \& {Venables}}]{Bruch2007}
{Bruch}, L.~W., {Diehl}, R.~D., \& {Venables}, J.~A. 2007, Reviews of Modern
  Physics, 79, 1381

\bibitem[{{Burke} \& {Brown}(2010)}]{Burke2010}
{Burke}, D.~J., \& {Brown}, W.~A. 2010, Physical Chemistry Chemical Physics
  (Incorporating Faraday Transactions), 12, 5947

\bibitem[{{Collings} {et~al.}(2004){Collings}, {Anderson}, {Chen}, {Dever},
  {Viti}, {Williams}, \& {McCoustra}}]{Collings2004}
{Collings}, M.~P., {Anderson}, M.~A., {Chen}, R., {et~al.} 2004, \mnras, 354,
  1133

\bibitem[{{Collings} {et~al.}(2015){Collings}, {Frankland}, {Lasne},
  {Marchione}, {Rosu-Finsen}, \& {McCoustra}}]{Collings2015}
{Collings}, M.~P., {Frankland}, V.~L., {Lasne}, J., {et~al.} 2015, \mnras, 449,
  1826

\bibitem[{{Edridge} {et~al.}(2013){Edridge}, {Freimann}, {Burke}, \&
  {Brown}}]{Edridge2013}
{Edridge}, J.~L., {Freimann}, K., {Burke}, D.~J., \& {Brown}, W.~A. 2013,
  Philosophical Transactions of the Royal Society of London Series A, 371,
  20110578

\bibitem[{{Garrod} {et~al.}(2008){Garrod}, {Weaver}, \& {Herbst}}]{Garrod2008}
{Garrod}, R.~T., {Weaver}, S.~L.~W., \& {Herbst}, E. 2008, \apj, 682, 283

\bibitem[{{Hama} \& {Watanabe}(2013)}]{Hama2013}
{Hama}, T., \& {Watanabe}, N. 2013, Chemical Reviews, 113, 8783

\bibitem[{{Hasegawa} \& {Herbst}(1993)}]{Hasegawa1993}
{Hasegawa}, T.~I., \& {Herbst}, E. 1993, \mnras, 261, 83

\bibitem[{{He} {et~al.}(2016){He}, Acharyya, \& {Vidali}}]{He2016a}
{He}, J., Acharyya, K., \& {Vidali}, G. 2016, ApJ submitted

\bibitem[{{He} {et~al.}(2015){He}, {Shi}, {Hopkins}, {Vidali}, \&
  {Kaufman}}]{He2015b}
{He}, J., {Shi}, J., {Hopkins}, T., {Vidali}, G., \& {Kaufman}, M.~J. 2015,
  \apj, 801, 120

\bibitem[{{He} \& {Vidali}(2014)}]{He2014a}
{He}, J., \& {Vidali}, G. 2014, Faraday Discussions, 168, 517

\bibitem[{{Ikeda} {et~al.}(2001){Ikeda}, {Ohishi}, {Nummelin}, {Dickens},
  {Bergman}, {Hjalmarson}, \& {Irvine}}]{Ikeda2001}
{Ikeda}, M., {Ohishi}, M., {Nummelin}, A., {et~al.} 2001, \apj, 560, 792

\bibitem[{{Jing} {et~al.}(2013){Jing}, {He}, {Bonini}, {Brucato}, \&
  {Vidali}}]{Jing2013}
{Jing}, D., {He}, J., {Bonini}, M., {Brucato}, J.~R., \& {Vidali}, G. 2013,
  Journal of Physical Chemistry A, 117, 3009

\bibitem[{{Karssemeijer} {et~al.}(2014){Karssemeijer}, {de Wijs}, \&
  {Cuppen}}]{Karssemeijer2014}
{Karssemeijer}, L.~J., {de Wijs}, G.~A., \& {Cuppen}, H.~M. 2014, Physical
  Chemistry Chemical Physics (Incorporating Faraday Transactions), 16, 15630

\bibitem[{{Katz} {et~al.}(1999){Katz}, {Furman}, {Biham}, {Pirronello}, \&
  {Vidali}}]{Katz1999}
{Katz}, N., {Furman}, I., {Biham}, O., {Pirronello}, V., \& {Vidali}, G. 1999,
  \apj, 522, 305

\bibitem[{Kolasinski(2008)}]{Kolasinski2008}
Kolasinski, K. 2008, Surface Science: Foundations of Catalysis and Nanoscience
  (Wiley)

\bibitem[{Linnartz {et~al.}(2015)Linnartz, Ioppolo, \& Fedoseev}]{Linnartz2015}
Linnartz, H., Ioppolo, S., \& Fedoseev, G. 2015, Int. Rev. Phys. Chem., 34, 205

\bibitem[{{Mart{\'{\i}}n-Dom{\'e}nech}
  {et~al.}(2014){Mart{\'{\i}}n-Dom{\'e}nech}, {Mu{\~n}oz Caro}, {Bueno}, \&
  {Goesmann}}]{Martin2014}
{Mart{\'{\i}}n-Dom{\'e}nech}, R., {Mu{\~n}oz Caro}, G.~M., {Bueno}, J., \&
  {Goesmann}, F. 2014, \aap, 564, A8

\bibitem[{{Noble} {et~al.}(2012){Noble}, {Congiu}, {Dulieu}, \&
  {Fraser}}]{Noble2012}
{Noble}, J.~A., {Congiu}, E., {Dulieu}, F., \& {Fraser}, H.~J. 2012, \mnras,
  421, 768

\bibitem[{{{\"O}berg} {et~al.}(2009){{\"O}berg}, {Fayolle}, {Cuppen}, {van
  Dishoeck}, \& {Linnartz}}]{Oberg2009}
{{\"O}berg}, K.~I., {Fayolle}, E.~C., {Cuppen}, H.~M., {van Dishoeck}, E.~F.,
  \& {Linnartz}, H. 2009, \aap, 505, 183

\bibitem[{{Pirronello} {et~al.}(1997){Pirronello}, {Biham}, {Liu}, {Shen}, \&
  {Vidali}}]{Pirronello1997b}
{Pirronello}, V., {Biham}, O., {Liu}, C., {Shen}, L., \& {Vidali}, G. 1997,
  \apjl, 483, L131

\bibitem[{Smith {et~al.}(2016)Smith, May, \& Kay}]{Smith2016}
Smith, R.~S., May, R.~A., \& Kay, B.~D. 2016, J. Phys. Chem. B, 120, 1979

\bibitem[{{Vasyunin} {et~al.}(2009){Vasyunin}, {Semenov}, {Wiebe}, \&
  {Henning}}]{Vasyunin2009}
{Vasyunin}, A.~I., {Semenov}, D.~A., {Wiebe}, D.~S., \& {Henning}, T. 2009,
  \apj, 691, 1459

\bibitem[{{Vidali}(2013)}]{Vidali2013}
{Vidali}, G. 2013, Chemical Reviews, 113, 8752

\end{thebibliography}
\bibliographystyle{apj}
\end{document}